\documentclass{ws-procs975x65}
\usepackage{graphicx}
\usepackage{amsmath}
\usepackage{array}

\begin{document}

\title{Gravitational phase transition of self-gravitating\\systems of fermions in General Relativity}
\author{Giuseppe ALBERTI$^*$}
\address{Living Systems Research,\\
Roseggerstra\ss e 27/2, A-9020 Klagenfurt am W\"{o}rthersee, Austria\\
$^*$E-mail: giuseppe.alberti@ilsr.at}

\author{Pierre-Henri CHAVANIS$^*$}
\address{Laboratoire de Physique Th\'eorique (UMR 5152), Universit\'e Paul Sabatier,\\
118 Route de Narbonne, F-31062 Toulouse, France\\
$^*$E-mail: chavanis@irsamc.ups-tlse.fr}

\begin{abstract}
The Thomas-Fermi model is extended at finite temperature, to describe the gravitational phase transition occurring in massive fermionic systems in a general-relativistic framework. It is shown that, when a nondegenerate fermionic gas (for $N < N_{OV}$, where $N_{OV}$ is the Oppenheimer-Volkoff limit) is cooled down below a critical temperature, a condensed phase emerges and the gravitational collapse is prevented by quantum mechanics. If $N > N_{OV}$, by contrast, the system is destined to collapse towards a Black Hole because no equilibrium states exist.
\end{abstract}


\bodymatter

\section{Introduction} \label{sec:1}

\noindent The object of this work, intendend as a r\'{e}sum\'{e} of a previous publication\cite{ac_2018}, is to describe the occurrence of the gravitational phase transition that the Fermi gas at non-zero temperature experiences, in a general relativistic framework.

\noindent As it is known, gravitational phase transitions have been investigated since early '70s, when Hertel \& Thirring \cite{ht_1971} have shown the non-equivalence of statistical ensembles. Several steps forward have been made by Padmanabhan \cite{padmanabhan_1990} and Chavanis \cite{chavanis_2002} (for a detailed review see Ref.\cite{chavanis_2006} and references therein) who have shown the occurrence of the phase transition in several systems (e.g. hard spheres,...).

\noindent In the case of fermions\cite{chavanis_2002}, when the system is cooled down below a critical temperature, a condensed phase emerges and the initial gaseous configuration evolves towards this condensed configuration.

\noindent Bili\'{c} \& Viollier \cite{bv_1999} firstly studied the occurrence of the gravitational phase transition in General Relativity (hereafter GR): however, they focused their attention only on a particular case. For this reason, in this work, we complete their previous investigations by describing the most general case.

\noindent The paper is organized as follows. In Sec. \ref{sec:2} we present the main equations. In Secs. \ref{sec:3} and \ref{sec:4} we discuss the occurrence of the phase transition in the canonical and microcanonical ensembles, respectively. In Sec. \ref{sec:5}, finally, we draw some conclusions.

\section{Theoretical framework} \label{sec:2}

\noindent We consider a (static) self-gravitating\footnote{In this work we neglect the contribution of the other interactions that, in a more realistic situation, fermions certainly feel.} fermionic gas, formed by \emph{N} particles of mass \emph{m} at a temperature $T \neq 0$ and placed within a spherical box of dimension \emph{R}. The equilibrium equations are given by

\begin{subequations} \label{eq:sistema_tov}
\begin{align}
& \frac{d\Phi}{dr} = -\frac{2G}{c^4} \frac{(\Phi + 1) (M_r c^2 + 4\pi Pr^3)}{r^2} \Biggl(1 - \frac{2GM_r}{rc^2}\Biggr)^{-1} \,, \label{subeq:sistema_tov_1} \\
& \frac{dM_r}{dr} = \frac{4\pi \epsilon r^2}{c^2} \,. \label{subeq:sistema_tov_2}
\end{align}
\end{subequations}

\noindent with the conditions $\Phi(0)=\Phi_0$ and $M_r(0)=0$. In the previous equations, $\Phi$ corresponds to the gravitational potential, $M_r$ represents the mass-energy contained within a sphere placed at distance $r < R$ from the center of the system ($\epsilon$ and \emph{P} are the mass-energy density and the pressure, respectively). The particle number \emph{N} is given by

\begin{equation} \label{eq:numero_barionico}
N = N(\Phi_0) = \int_0^R 4\pi n r^2 \Bigl(1 - \frac{2GM_r}{rc^2}\Bigr)^{-1/2} dr \,,
\end{equation}

\noindent being $n = \rho/m$ the particle number density ($\rho$ is the rest mass density). The thermodynamic analysis is performed by means of the caloric curve $T = T(E)$\footnote{For the details concerning the numerical procedure used to get the caloric curve see Ref.\cite{ac_2018}.}. To this purpose, we define the following variables

\begin{equation} \label{eq:variabili_caloriche}
\Lambda = -\frac{E_bR}{GN^2m^2} = \frac{(Nm - M)Rc^2}{GN^2m^2} \,, \qquad \eta = \frac{GNm^2}{k_BTR} \,.
\end{equation}

\section{Canonical instabilities} \label{sec:3}

\noindent In this Section we discuss the case of the canonical ensemble. In Fig. 1 we have represented the occurrence of the phase transition for $R = 15\, R_{OV}$ and $N = 0.7277\, N_{OV}$ (left panel) and $R = 15\, R_{OV}$ and $N = 1.0012\, N_{OV}$\footnote{$R_{OV}$ and $N_{OV}$ are the values of the radius and of the particle number at the Oppenheimer-Volkoff limit \cite{ov_1939}. Numerically we have $R_{OV} = 9.162$ km and $N_{OV} = 8.752 \times 10^{56}$ particles.} (right panel).

\noindent In both cases the system exhibits a phase transition because we observe the coexistence, at the same (transition) temperature ($\eta_t = 1.5722$ in the first case and $\eta_t = 1.3315$ in the second one), of several states identified by the points $P_1$, $P_2$, $P_3$ and $P_4$. The plots present different colors according to the stability of the solutions (black and green lines corresponding to stable states, red lines corresponding to unstable states). Let us consider the case $N < N_{OV}$ first.

\begin{figure} \label{fig:canonico}
\begin{center}
\includegraphics[scale = 0.25]{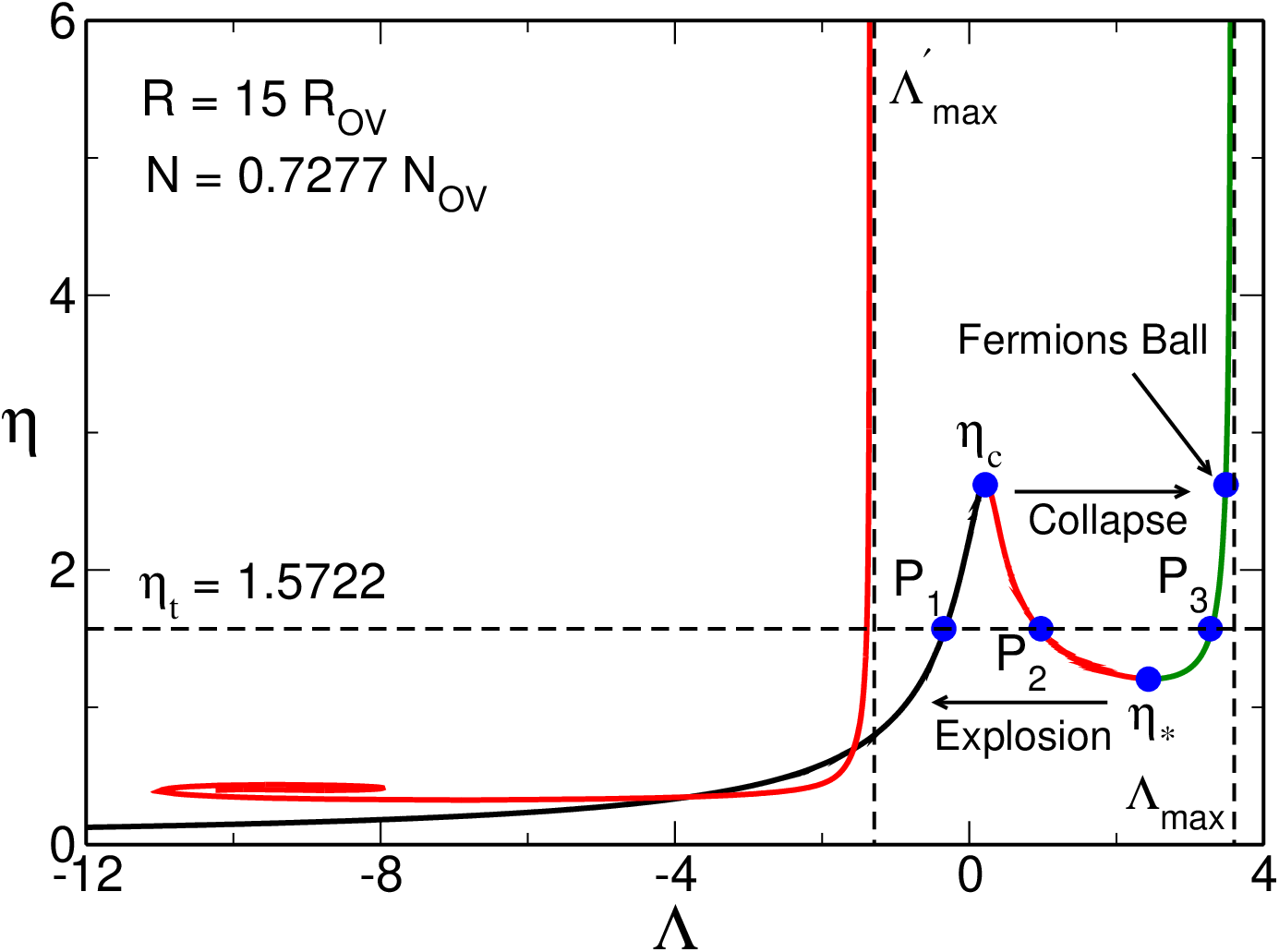}
\includegraphics[scale = 0.25]{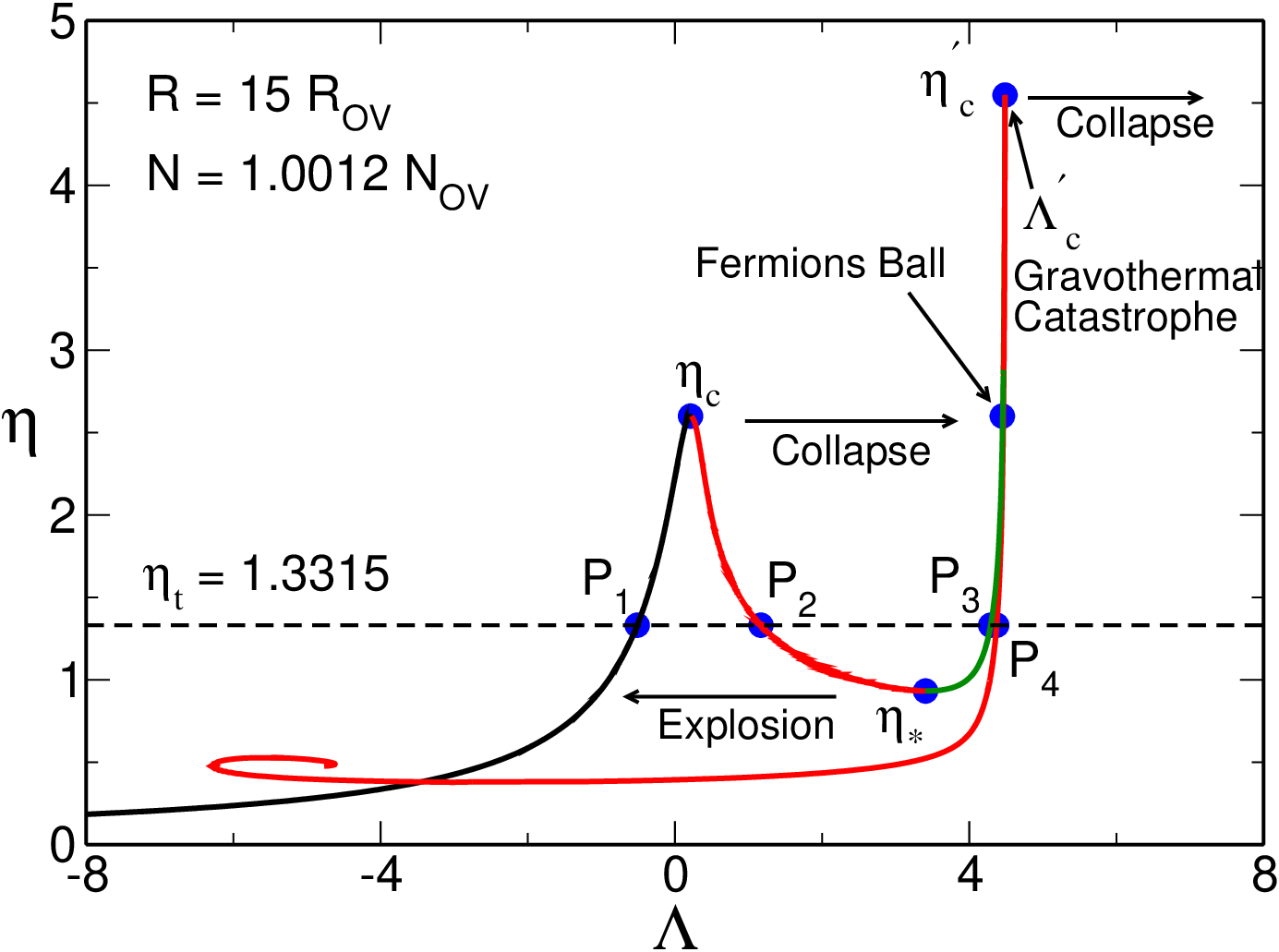}
\caption{Equilibrium phase diagrams for fermionic systems in GR. $\eta_t$ corresponds to the transition temperature. \textbf{Left Panel}: At the temperature $\eta_c$ the system evolves from the gaseous (black line) to the condensed phase (green line). The gravitational collapse is prevented by Pauli's exclusion principle. At the temperature $\eta_*$, the system evolves from the condensed phase to the gaseous one. The gravitational explosion is halted by the box. \textbf{Right Panel}: The condensed phase collapses towards a Black Hole at the temperature $\eta'_c$.}
\end{center}
\end{figure}

\begin{figure} \label{fig:energia_libera}
\begin{center}
\includegraphics[scale = 0.25]{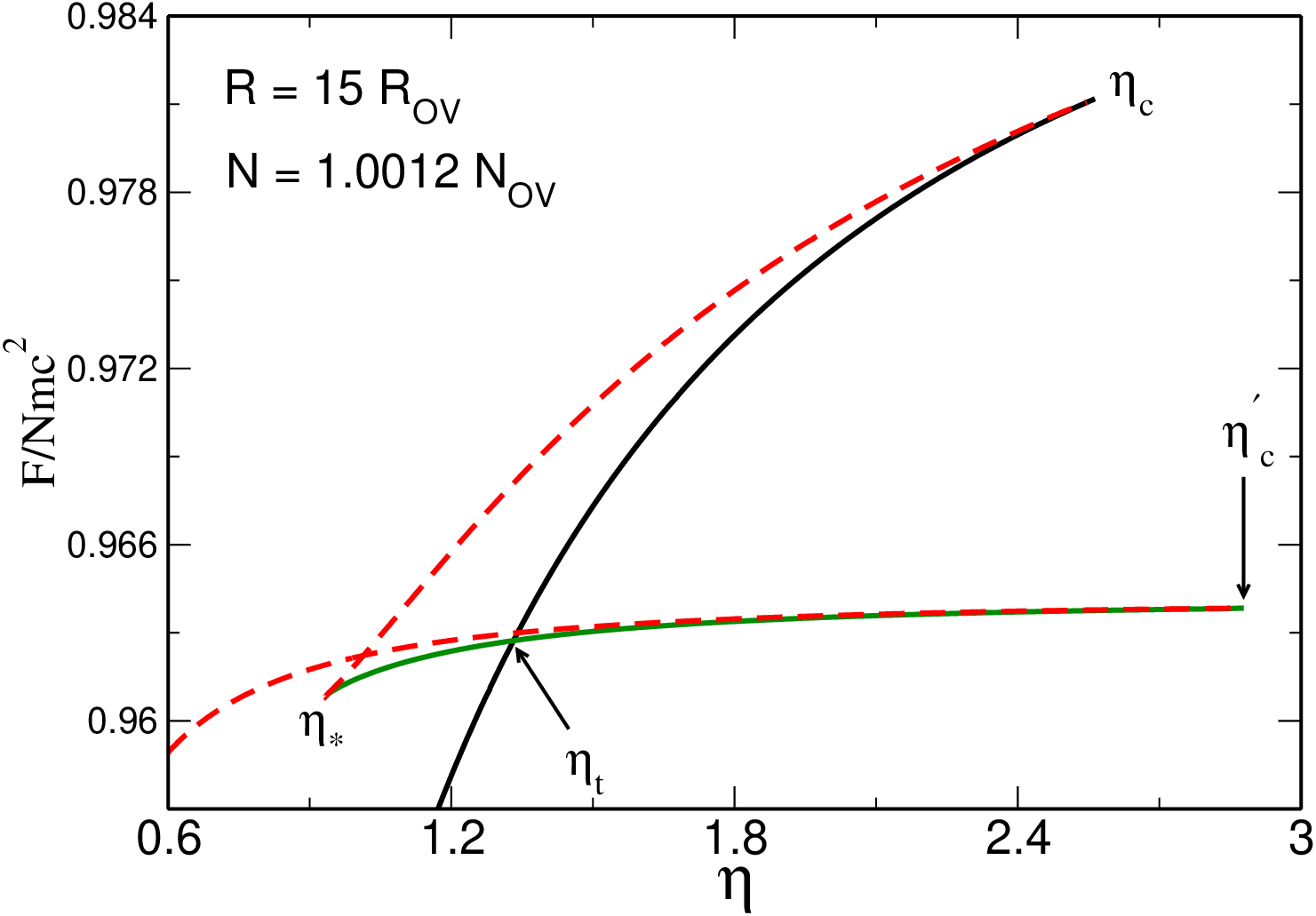}
\includegraphics[scale = 0.25]{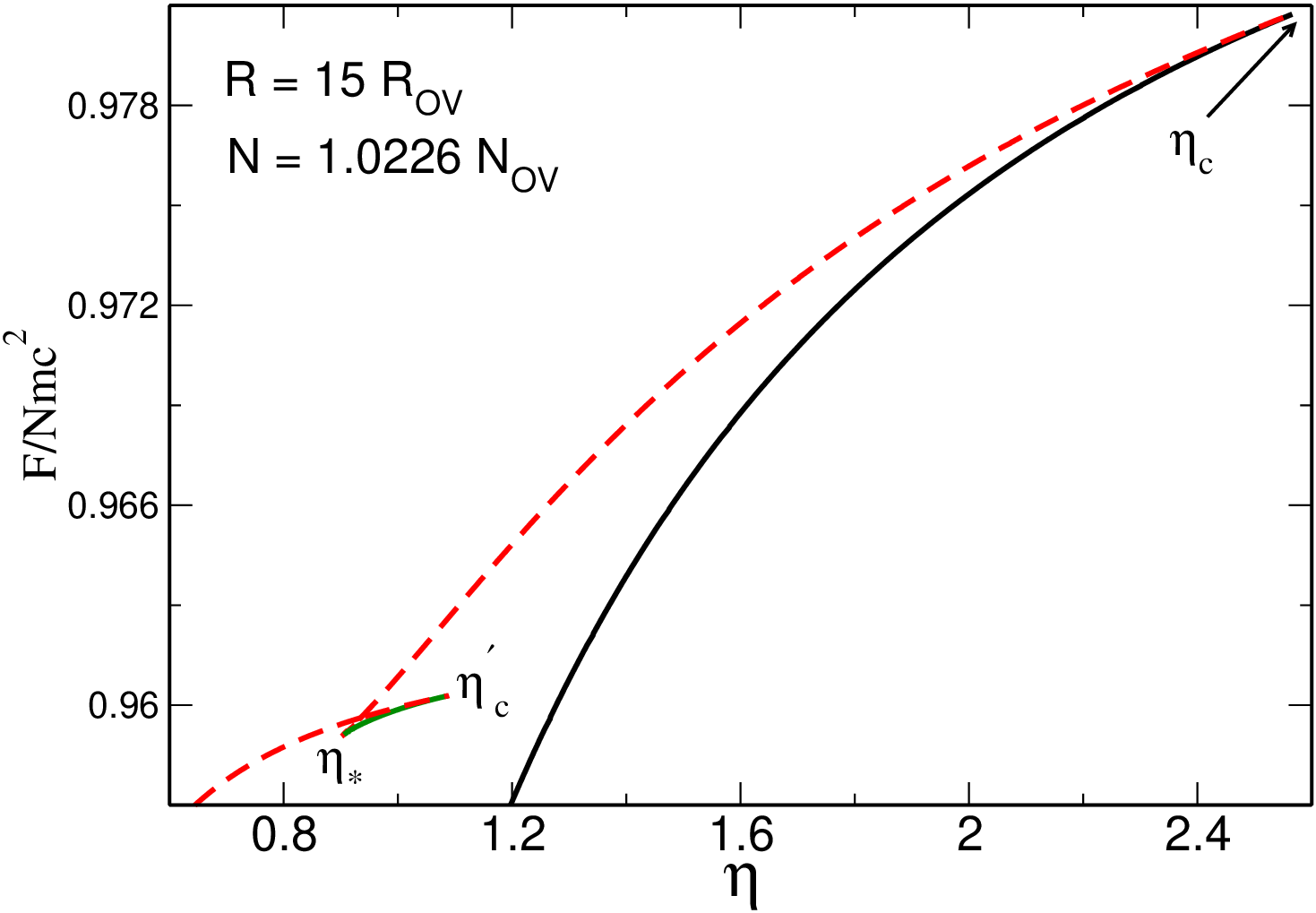}
\caption{Free energy as a function of the normalized inverse temperature $\eta$. \textbf{Left Panel}: The phase transition is identified by the intersection point between the gaseous (black full line) and the condensed branch (green full line). \textbf{Right Panel}: The intersection point between the gaseous and the condensed phase is missing: the phase transition is thus suppressed (the crossing point concerns unstable states).}
\end{center}
\end{figure}

\noindent Points $P_1$ represent the gaseous phase (black line): they are global minima of free energy (thus stable) for $\eta < \eta_t$ and local minima of free energy (thus metastable\footnote{The lifetime of metastable states can be longer than the age of the Universe. This implies that the phase transition does not occur in practice. However, metastable states can play a central role in astrophysics\cite{chavanis_2006}.}) for $\eta > \eta_t$. When the temperature exceeds the critical value $\eta_c$, the gaseous phase undergoes a collapse and forms a compact object (condensed phase, green line, points $P_3$) containing all the mass. The stability of the solutions $P_3$ is inverted with respect to that of the solutions $P_1$. Consequently, the compact object represents a stable configuration (it is a global minimum of free energy). We refer to this compact object as ``fermion ball".

\noindent The condensed phase, similar to the gaseous one, evolves too. For $\eta < \eta_*$, indeed, the condensed undergoes an explosion (halted by the box). Points $P_2$, by contrast, correspond to unstable physical solutions (they are saddle points). Moreover, in this region of the diagram, the specific heat is negative (because $d\eta/d\Lambda < 0$).

\noindent Let us now turn our attention to the case $N > N_{OV}$ (right panel of Fig. 1). The difference, with respect to the former case, is represented by the presence of a second collapse (of general relativistic origin) at the temperature $\eta'_c$. In this case, the condensed phase undergoes a second collapse towards a Black Hole.

\noindent The temperature $\eta'_c$, which is a decreasing function of \emph{N}, allows un upper limit for the extension of the condensed phase. The theory identifies a critical value of \emph{N}, namely that corresponding to the extinction of the condensed phase and to the suppression of the phase transition\cite{ac_2018}.

\noindent The reason of this phenomenon is shown in Fig. 2. The left panel plots the free energy as a function of the (reverse) normalized temperature, for $N = 1.0012 \, N_{OV}$. The phase transition is identified by the intersection point between the gaseous and the condensed phase. The right panel shows the case $N = 1.0226 \, N_{OV}$. As we see, the plot does not display any intersection point between the two phases, implying that the phase transition is suppressed.

\section{Microcanonical instabilities} \label{sec:4}

\noindent Let us now discuss the microcanonical ensemble. In Fig. 3 we study the occurrence of the phase transition for $R = 179\, R_{OV}$ and $N = 3.2620\, N_{OV}$ (left panel) and $R = 179\, R_{OV}$ and $N = 4.2657\, N_{OV}$ (right panel). Similar to the canonical ensemble, the left panel of Fig. 3 shows the occurrence of the phase transition because of the coexistence, at the same (transition) energy $\Lambda_t = -0.025$, of several states identified by the points $P_1$, $P_2$, $P_3$ and $P_4$.

\noindent Points $P_1$ represent the gaseous phase (black line): they are global entropy maxima (thus stable) for $\Lambda < \Lambda_t$ and local entropy maxima (thus metastable) for $\Lambda > \Lambda_t$. When the energy exceeds the critical value $\Lambda_c$, the gaseous phase undergoes a collapse and forms a compact object (condensed phase, green line, points $P_3$) containing a fraction ($\sim 1/4$) of the total mass and surrounded by an atmosphere.

\noindent The stability of the solutions $P_3$ is inverted with respect to that of the solutions $P_1$. Consequently, the compact object represents a stable configuration (it is a global entropy maximum). We refer to this compact object as ``fermion ball".

\noindent The condensed phase, similar to the gaseous one, evolves too. For $\Lambda < \Lambda_*$, it undergoes an explosion (halted by the box). Analogous to the canonical ensemble, points $P_2$ are unstable saddle points. In this region of the diagram the specific heat is negative (because $d\eta/d\Lambda < 0$).

\begin{figure} \label{fig:microcanonico}
\begin{center}
\includegraphics[scale = 0.25]{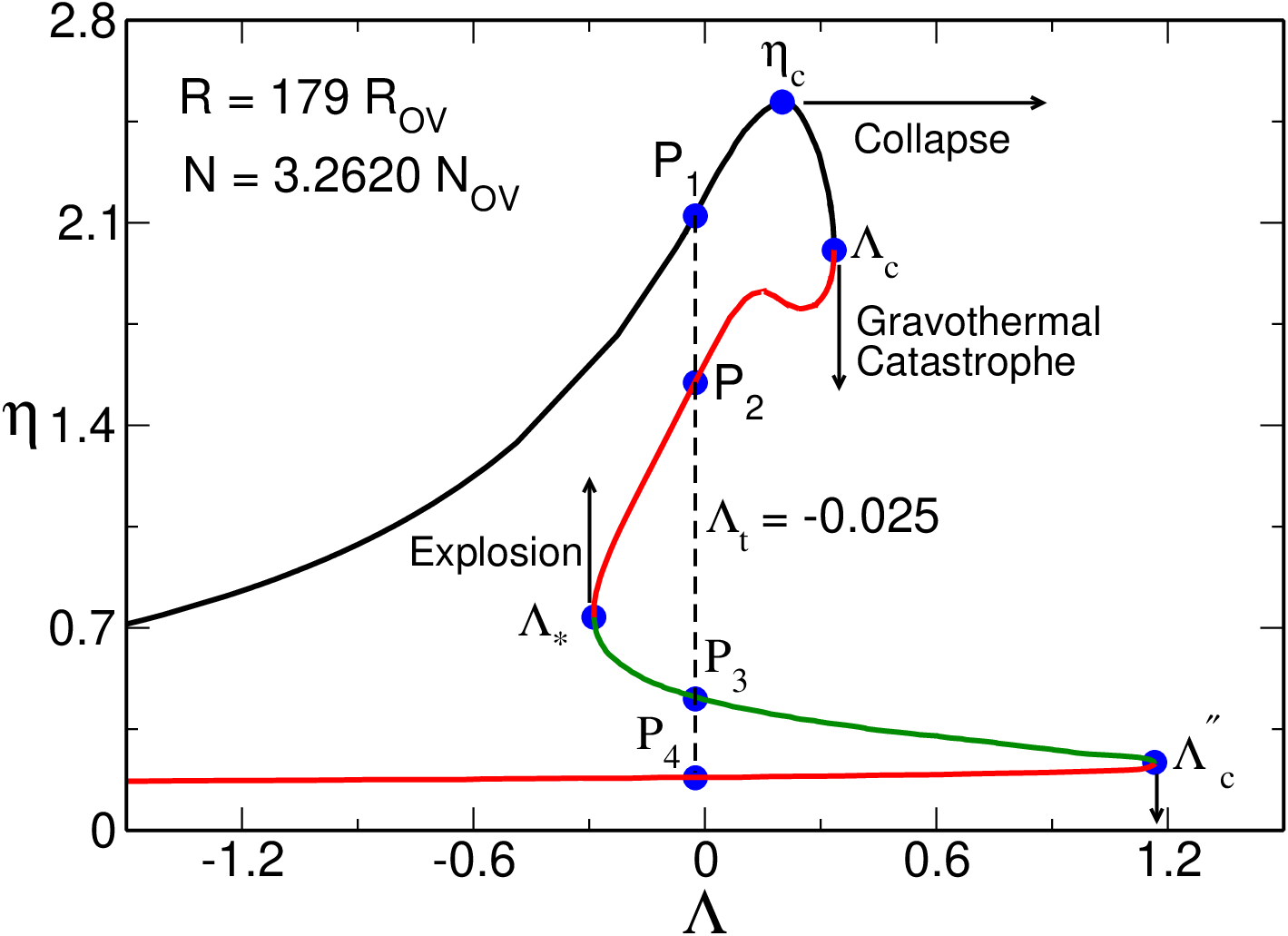}
\includegraphics[scale = 0.25]{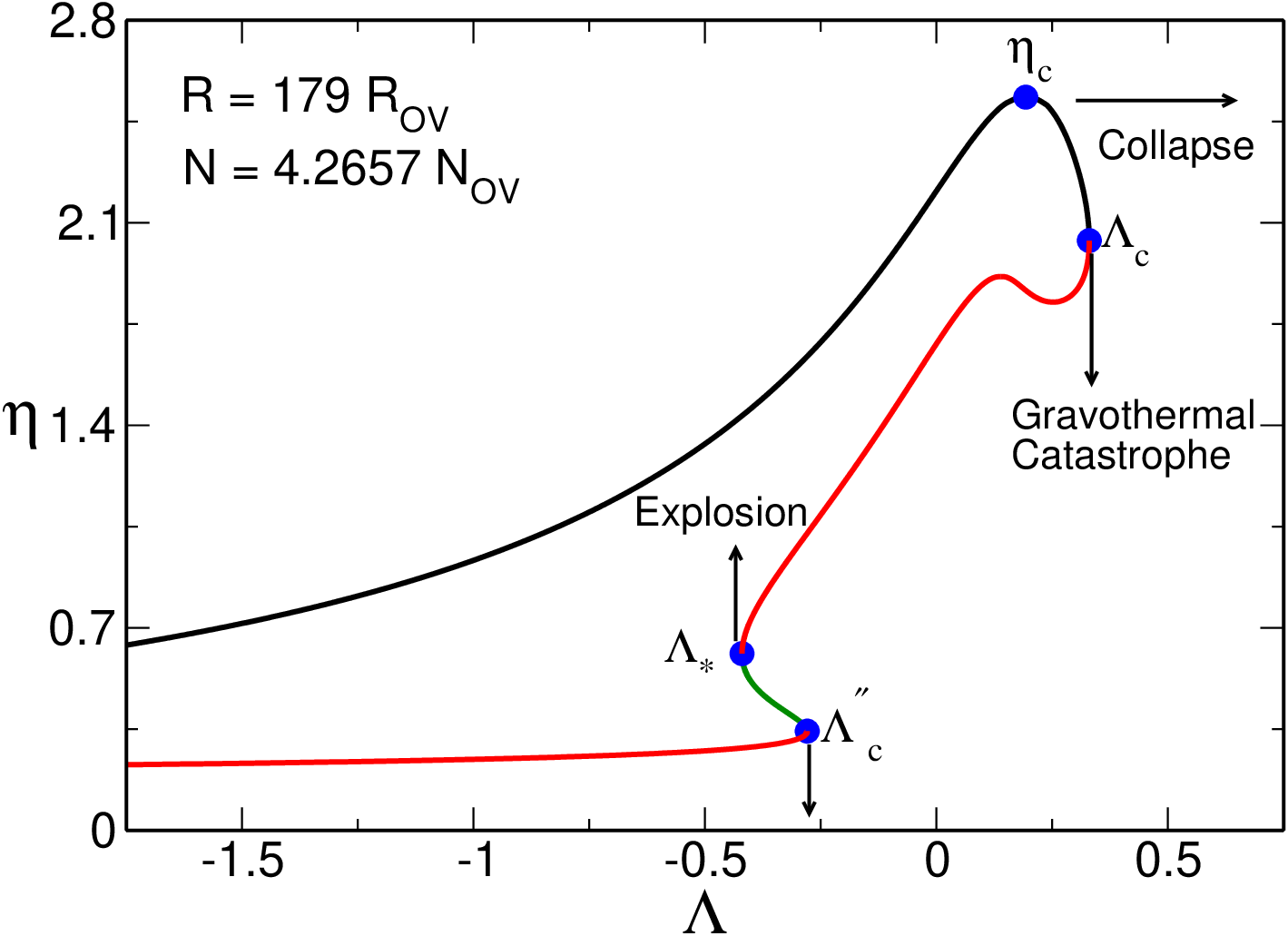}
\caption{Equilibrium phase diagrams for fermionic systems in GR. \textbf{Left Panel}: A phase transition from the gaseous to the condensed phase occurs at the energy $\Lambda_t$. At the energy $\Lambda_c$ the system evolves from the gaseous to the condensed phase. The gravitational collapse is prevented by quantum degeneracy. At the energy $\Lambda_*$ the system evolves from the condensed to the gaseous phase. The gravitational explosion is halted by the box. At the energy $\Lambda''_c$ the system collapses towards a Black Hole. \textbf{Right Panel}: The phase transition is suppressed.}
\end{center}
\end{figure}

\begin{figure} \label{fig:entropia}
\begin{center}
\includegraphics[scale = 0.24]{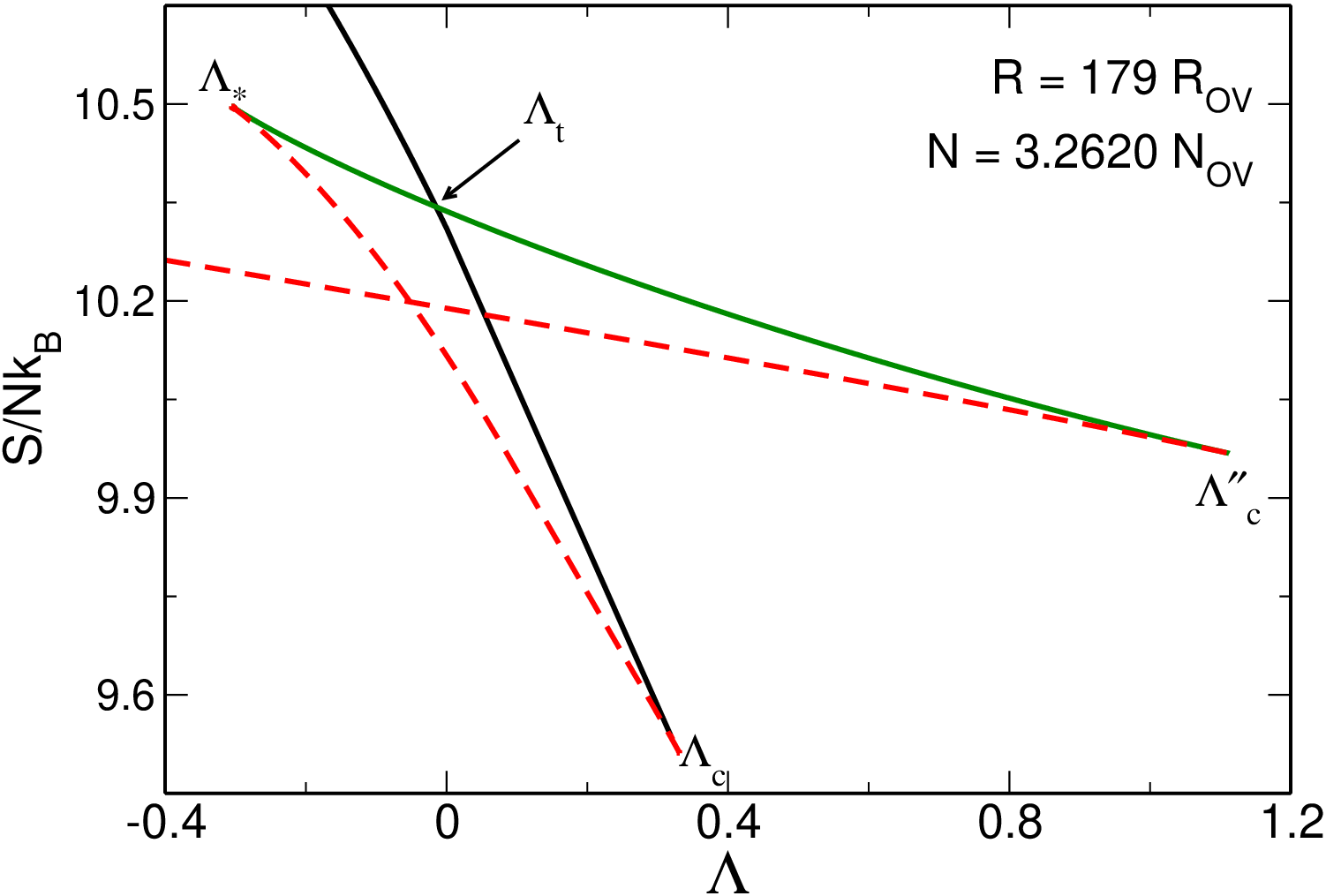}
\includegraphics[scale = 0.24]{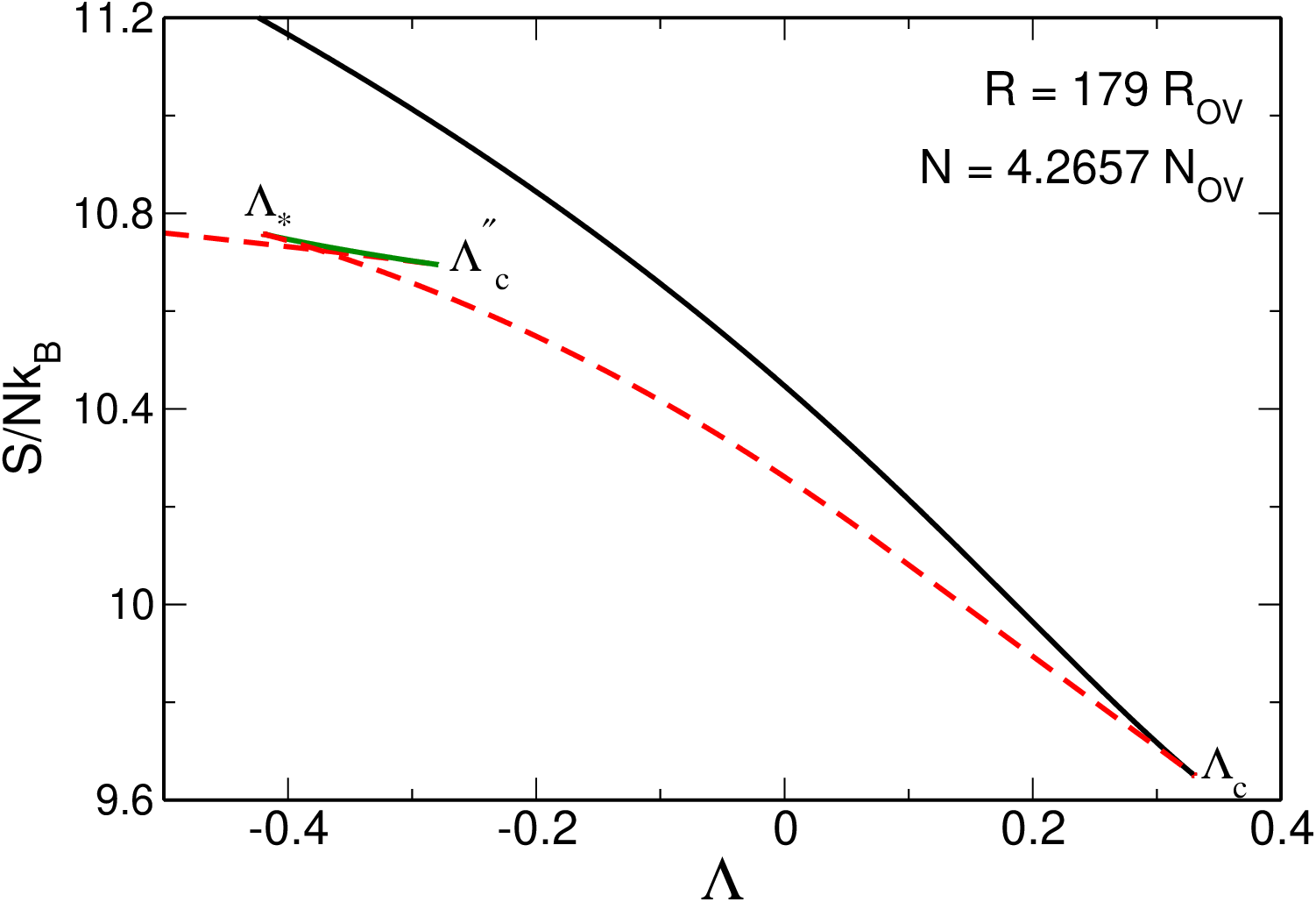}
\caption{Entropy as a function of the normalized energy $\Lambda$. \textbf{Left Panel}: The phase transition is identified by the intersection point between the gaseous (black full line) and the condensed branch (green full line). \textbf{Right Panel}: The intersection point between the gaseous and the condensed phase is missing, so the phase transition is suppressed (the crossing point concerns unstable states).}
\end{center}
\end{figure}

\noindent The novelty with respect to the non-relativistic regime\cite{chavanis_2002} is the presence of a second collapse, occurring at the critical energy $\Lambda''_c$. The existence of this critical energy allows an upper limit to the extension of the condensed phase and, as a consequence, the suppression of the phase transition\cite{ac_2018}.

\noindent To better understand this phenomenon we have represented, in Fig. 4, the entropy as a function of the normalized energy $\Lambda$. The left panel shows the occurrence of the phase transition, because of the presence of a crossing point between the gaseous and the condensed phase. The right panel of Fig. 4, by contrast, does not display any crossing point between the two phases and the phase transition is thus suppressed (Fig. 3 plots the caloric curve associated).

\section{Concluding remarks} \label{sec:5}

\noindent In this work we have studied the nature of phase transitions of the Fermi gas in a general relativistic framework. The model takes both quantum mechanics and GR into account. As we have seen, the occurrence of the phase transition depends on the values of \emph{N} and \emph{R}.

\noindent In the main paper\cite{ac_2018} we have determined two critical values of the cavity radius, namely the canonical critical radius $R_{CCP} = 3.57\, R_{OV}$ and the microcanonical critical radius $R_{MCP} = 27.4\, R_{OV}$. If $R < R_{CCP}$, the system does not experience any phase transition whereas, if $R_{CCP} \leq R < R_{MCP}$, the system experiences the canonical phase transition. If $R \geq R_{MCP}$, both types of phase transition occur.

\noindent In Sec. \ref{sec:3} we have considered the case of the canonical phase transition. We have seen that, for $N < N_{OV}$, the result is similar to that obtained in the non-relativistic regime\cite{chavanis_2002}. The result of the phase transition is, indeed, the formation of a compact object containing all the mass of the initial configuration. Things change when we consider the case $N \geq N_{OV}$, because the system exhibits a second collapse at the temperature $\eta'_c$. The result of this collapse is a Black Hole. However, if the value of \emph{N} is above a critical one (see Ref.\cite{ac_2018}), the phase transition is suppressed.

\noindent In Sec. \ref{sec:4} we have considered the case of the microcanonical phase transition. Similar to the canonical ensemble, the system exhibits a second collapse (at the energy $\Lambda''_c$) towards a Black Hole. Analogous to the canonical ensemble, we observe the phenomenon of the suppression of the phase transition.

\noindent The results obtained in this work call for further investigations. For example, the microcanonial phase transitions may be related to the onset of red-giant structure or to the Supernova phenomenon. Furthermore, to check the robustness of the model and have a first ``experimental" proof of the occurrence of the phase transition, it would be interesting to perform \emph{N}-body simulations in GR. This could represent a new challenge for Numerical Relativity.

\end{document}